\documentclass[a4paper]{article}

\usepackage[T1]{fontenc}
\usepackage[utf8]{inputenc}

\usepackage[tt=false]{libertine}
\usepackage{libertinust1math}

\usepackage{hyperref}

\usepackage{graphicx}

\usepackage[style=authoryear-comp]{biblatex}
\title{The periodic zeta covariance function for Gaussian process regression}

\author{Giacomo Petrillo\\University of Florence\\Department of Statistics,
Computer Science and Applications (DiSIA)}

\begin{document}
    
    \renewcommand{\sectionautorefname}{Section}
    
    \maketitle
    
    \begin{abstract}
        
        I consider the Lerch-Hurwitz or periodic zeta function as covariance
        function of a periodic continuous-time stationary stochastic process.
        The function can be parametrized with a continuous index $\nu$ which
        regulates the continuity and differentiability properties of the
        process in a way completely analogous to the parameter $\nu$ of the
        Matérn class of covariance functions. This makes the periodic zeta a
        good companion to add a power-law prior spectrum seasonal component to
        a Matérn prior for Gaussian process regression. It is also a close
        relative of the circular Matérn covariance, and likewise can be used on
        spheres up to dimension three. Since this special function is not
        generally available in standard libraries, I explain in detail the
        numerical implementation.
        
    \end{abstract}
    
    \section{Introduction}
    \label{sec:intro}
    
    A Gaussian process, or Gaussian random field, is a distribution over an
    infinite dimensional space with Normal finite marginals; a
    finite-dimensional multivariate Normal is considered a specific case.
    Gaussian processes are used as distributions for statistical inference on
    functions with domain in time, space, or in general over quantities which
    have no fixed finite enumeration, that do not have a strongly constrained
    form. Applications range from geostatistics to optimization and machine
    learning. As general references, consider \cite{stein1999},
    \cite{wendland2004}, \cite{rasmussen2006}, and \cite{gramacy2020}.
    
    A Gaussian process is characterized by its covariance function, a
    two-point function that produces the entries of any marginal covariance
    matrix. In \S6.7, \cite{stein1999} considers a covariance function with
    spectral mass over the lattice $\mathbf j \in \mathbb Z^d$ of the kind
    $(\alpha^2 + |\mathbf j|^2)^{-(\nu + d/2)}$, with parameters $\alpha, \nu >
    0$ (see also \cite[eq.~4]{stein2005}), and proceeds to use it for inference
    with data on a regular square lattice using the discrete Fourier transform.
    In this article I calculate explicitly the covariance function for the $d =
    1$, $\alpha = 0$ case, such that it can be used for any design layout and
    combined arbitrarily with other covariance functions, which amounts to
    evaluating the periodic zeta function $F(x, s)$
    (\cite[p.~257]{apostol1976}).
    
    This periodic covariance is one of the many generalizations of the Matérn
    class of covariance functions (\cite[406]{handcock1993},
    \cite[p.~31]{stein1999}, \cite[p.~84]{rasmussen2006}), which include, to
    name a few examples, a non-stationary version (\cite[487]{paciorek2006}), a
    compactly supported version (\cite[eq.~10]{bevilacqua2022}), and a smooth
    extension to spheres (\cite[eq.~4]{jeong2015}). In particular,
    \cite{guinness2016} introduce an extension to spheres up to dimension
    three, naming it the circular Matérn covariance function, which amounts to
    considering the one-dimensional case of Stein's periodic Matérn and
    evaluating it on the great arc distance. (See also \cite{huang2022} for
    calculations.)
    
    \cite{guinness2016} compare the circular Matérn with many other
    alternatives on a pair of examples. Although it performs well, on a
    practical note they recommend using the chordal Matérn, i.e., the usual
    Matérn evaluated on the embedding of the sphere, due to the complications
    in computing the circular Matérn: they give a closed form solution only for
    half-integer $\nu$, and no quickly converging approximation scheme.
    \cite[365]{porcu2018}, mention this as an open problem, and
    \cite{alegria2021} propose as solution another covariance function with
    similar properties, the ``F-family'', defined in terms of the Gauss
    hypergeometric function. Here I show how to compute exactly a function very
    similar to the circular Matérn, providing yet another alternative, albeit
    only up to the 3-sphere.
        
    The layout of the article is as follows: \autoref{sec:pz} introduces the
    covariance function from first principles. \autoref{sec:num} describes how
    to compute it efficiently and accurately. Finally, \autoref{sec:concl}
    concludes by mentioning some possible extensions.
    
    \section{The periodic zeta covariance function}
    \label{sec:pz}
    
\begin{figure}
        \centerline{\includegraphics[width=80ex]{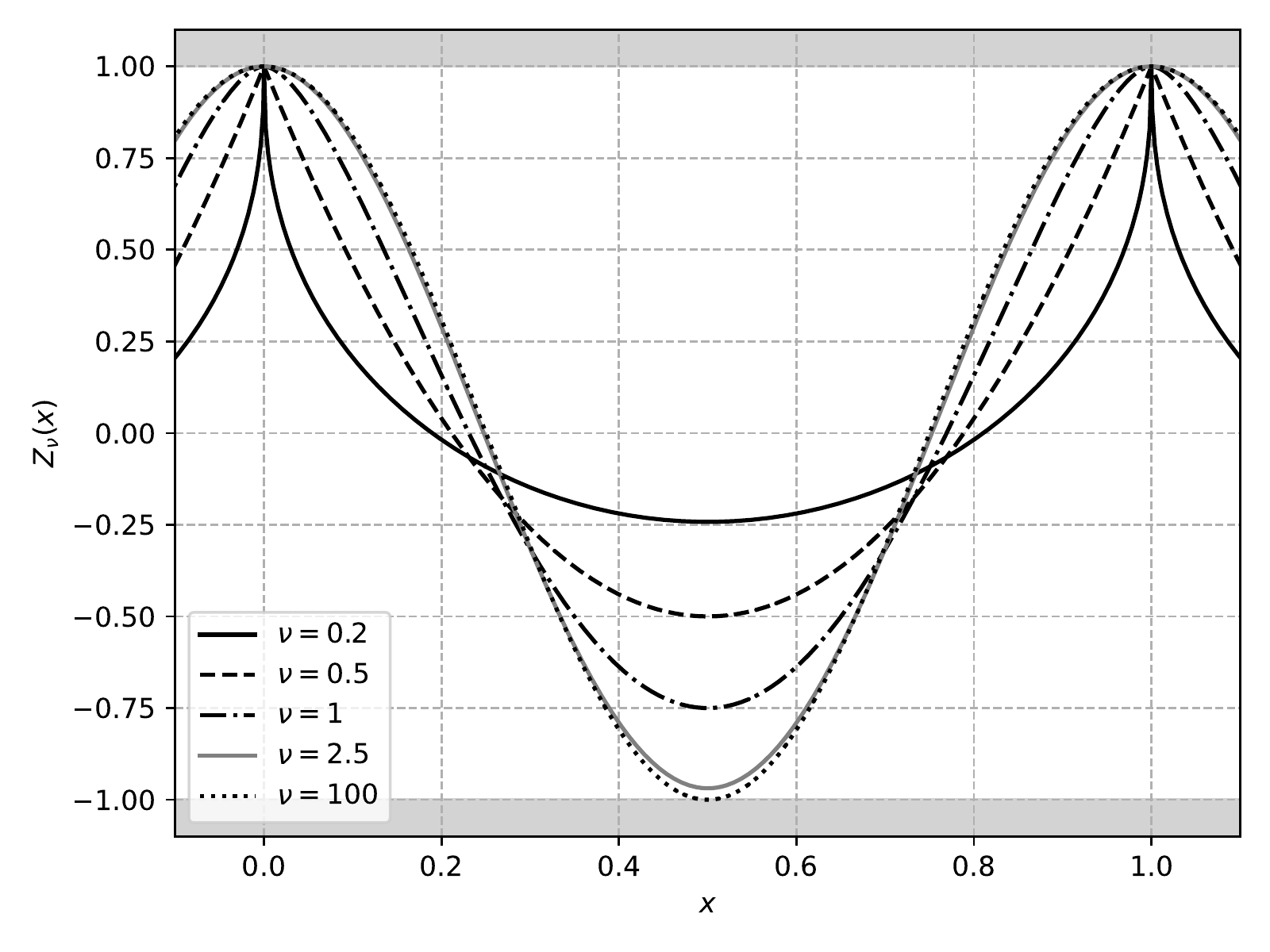}}
        \caption{\label{fig:znu} Plot of the covariance function $Z_\nu(x)$
        from \autoref{eq:znu} for some values of $\nu$. Due to
        property~\ref{it:sinf}, for large $\nu$ it becomes a cosine.}
    \end{figure}

    Consider the standard Fourier series basis of functions
    \begin{equation}
        \{ x \mapsto \cos(2\pi n x) \mid n = 0, 1, 2, \ldots \}
        \cup
        \{ x \mapsto \sin(2\pi n x) \mid n = 1, 2, 3, \ldots \},
    \end{equation}
    complete and orthonormal on the interval $x \in [0, 1]$. Let $f(x)$ be a
    stochastic process defined in terms of the distribution of its coefficients
    in the Fourier basis, without the intercept term:
    \begin{equation}
        f(x) = \sum_{n=1}^\infty \big(
            c_n \cos(2\pi n x) +
            s_n \sin(2\pi n x)
        \big),
    \end{equation}
    where the $c_n$ and $s_n$ are independently Normally distributed with
    variance
    \begin{equation}
        \operatorname{Var}[c_n] = \operatorname{Var}[s_n] = \frac 1 {n^s},
    \end{equation}
    for some $s > 1$. Since $f(x)$ is a linear combination of the coefficients,
    it is itself Normally distributed, i.e., a Gaussian process, with
    covariance function
    \begin{align}
        \operatorname{Cov}[f(x_1), f(x_2)]
        &= \sum_{n=1}^\infty \big(
            \operatorname{Var}[c_n] \cos(2\pi n x_1) \cos(2\pi n x_2) + {}
            \notag \\
        &\hphantom{{}=\sum_{n=1}^\infty \big(}
            \operatorname{Var}[s_n] \sin(2\pi n x_1) \sin(2\pi n x_2)
        \big) = \notag \\
        &= \sum_{n=1}^\infty \frac {\cos(2\pi n (x_1 - x_2))} {n^s}.
    \end{align}
    In this series we recognize the real part, evaluated at $x = x_1-x_2$, of
    the Lerch-Hurwitz or periodic zeta function
    \begin{equation}
        F(x, s) = E_s(x) =
        \sum_{n=1}^\infty \frac {e^{2\pi inx}} {n^s}. \label{eq:pzdef}
    \end{equation}
    The $F(x, s)$ notation is from the DLMF (\cite[\S25.13]{dlmf}), while
    $E_s(x)$ is from \cite[\S5.3]{crandall2012}. To summarize, we have that a
    Gaussian process which is diagonal in the Fourier basis with period~1 with
    a power-law spectrum has the periodic zeta function as covariance function.
    The properties of this function are:
    \begin{enumerate}
        \item It depends only on the distance $|x_1-x_2|$, so the process is
        stationary.
        \item The covariance function (and thus the process) is periodic with
        period~1.
        \item $F(x, s)$ converges absolutely for all $x$ if $s > 1$, and
        conditionally for non-integer $x$ if $s > 0$. \label{it:pzconv}
        \item $F(0, s) = \zeta(s)$, where $\zeta$ is Riemann's zeta function.
        \label{it:norm}
        \item $\partial_x F(x, s) = 2\pi i F(x, s - 1)$ for $s > 1$ and
        non-integer $x$. \label{it:deriv}
        \item $\lim_{s\to\infty} F(x, s) = e^{2\pi ix}$. \label{it:sinf}
    \end{enumerate}
    For the covariance function, I introduce the notation
    \begin{equation}
        Z_\nu(x) = \frac {\Re F(x, 1 + 2\nu)} {\zeta(1 + 2\nu)},
        \qquad \nu \ge 0, \label{eq:znu}
    \end{equation}
    with $x$ the difference $x_1 - x_2$, where for $\nu = 0$ I intend the
    limiting form of periodic white noise
    \begin{equation}
        Z_0(x) = \begin{cases}
            1 & x \bmod 1 = 0, \\
            0 & x \bmod 1 \neq 0.
        \end{cases}
    \end{equation}
    $Z_\nu$ may be called ``periodic zeta covariance function.''
    \autoref{fig:znu} shows it for some values of $\nu$. Due to
    property~\ref{it:norm}, it has unit variance. Due to~\ref{it:deriv}, the
    derivative of the process has covariance function
    \begin{align}
        \operatorname{Cov}[f'(x_1), f'(x_2)]
        &= \partial_{x_1} \partial_{x_2} Z_\nu(x_1 - x_2) = \notag \\
        &= (2\pi)^2 \frac {\zeta(1 + 2(\nu - 1))} {\zeta(1 + 2\nu)}
        Z_{\nu - 1}(x_1 - x_2). \label{eq:deriv}
    \end{align}
    For $s < 2$, $\partial_x \Re F(x, s)$ diverges for $x \to 0$, which means
    that, for $\nu < 1/2$, $Z_\nu(x)$ has a cusp in $x = 0$, indicating that
    the process is not mean-square Lipschitz-continuous. Together with
    \autoref{eq:deriv}, this implies that a process with covariance function
    $Z_\nu(x)$ is $\lceil\nu\rceil - 1$ times mean-square differentiable, and
    its highest order derivative is Lipschitz-continuous iff $\nu \bmod 1 \ge
    1/2$. See \autoref{sec:proofs} for details.
    
    Note that these properties w.r.t.\ $\nu$ are the same of the Matérn class
    of covariance functions (\cite[p.~84]{rasmussen2006},
    \cite[p.~31]{stein1999})
    \begin{equation}
        M_\nu(x) = \frac{2}{\Gamma(\nu)}
        \left(\frac r2\right)^\nu K_\nu(r),
        \qquad r = \sqrt{2\nu} x, \label{eq:matern}
    \end{equation}
    including the white noise limit
    \begin{equation}
        \lim_{\nu\to0^+} M_\nu(x) = \begin{cases}
            1 & x = 0, \\
            0 & x \ne 0.
        \end{cases}
    \end{equation}
    
    The circular Matérn covariance function (\cite[eq.~7]{guinness2016})
    \begin{equation}
        \psi_{\nu,\alpha}(\theta) \propto \sum_{k\in\mathbb Z}
        \frac {e^{ik\theta}} {(\alpha^2 + k^2)^{\nu+1/2}},
        \qquad \theta \in [0,\pi],
    \end{equation}
    is equivalent to the periodic zeta in the limit $\alpha\to 0^+$ with the
    divergent $k = 0$ intercept term removed and $\theta$ identified with $2\pi
    x$. Due to \cite[corollary~4a]{gneiting2013}, this is a positive definite
    function on the three-dimensional sphere $\mathbb S^3$ (and thus also
    $\mathbb S^2$) with $\theta$ the great arc distance. The same criterion
    also applies to the periodic zeta, in that $a/2^{s+1} + F(\theta/(2\pi),
    s)$ is a valid covariance function on $\mathbb S^3$ for $a \ge 1$ (see
    \autoref{sec:proofs}).
    
    What $Z_\nu$ lacks in comparison with $\psi_{\nu,\alpha}$ is a parameter
    regulating the correlation length like $\alpha$ does. To this end, consider
    the Lerch zeta function (\cite[p.~17]{laurincikas2002},
    \cite[\S25.14]{dlmf})
    \begin{equation}
        L(\lambda, \alpha, s) = \sum_{m=0}^\infty
        \frac {e^{2\pi i\lambda m}} {(m + \alpha)^s}, \label{eq:lerch}
    \end{equation}
    which counts as special case $F(x, s) = e^{2\pi ix} L(x, 1,
    s)$.\footnote{The Lerch function is also connected to the F-family of
    \cite{alegria2021} through \cite[eq.~9.559(1)]{gradshtein2014} for $\tau=1,
    \nu\to 0^+$, although it's not an interesting case.} Thus I define a
    ``Lerch covariance function''
    \begin{equation}
        Z_{\nu,\alpha}(x) = \frac
        {\Re e^{2\pi ix} L(x, \alpha, 1 + 2\nu)}
        {\zeta(1 + 2\nu, \alpha)},
        \qquad Z_{\nu,1}(x) = Z_\nu(x),
    \end{equation}
    which, like $Z_\nu$, has similar properties to the Matérn. Larger values of
    $\alpha$ produce a flatter spectrum head and thus shorten the correlation
    length. Although I do not treat in detail this extension, in
    \autoref{sec:concl} I give indications on how to compute it.
    
    \section{Numerical implementation}
    \label{sec:num}
    
    For large enough $s$, the series defining $F(x, s)$ (\autoref{eq:pzdef})
    converges rapidly and can be summed directly. In practice this is convenient
    for $s \gtrsim 10$.
    
    For small $s$, I make use of the relation with the Hurwitz zeta function
    (\cite[eq.~25.13.2]{dlmf})
    \begin{equation}
        F(x, s) = \frac{\Gamma(1 - s)}{(2\pi)^{1-s}} \left(
            e^{\pi i(1-s)/2} \zeta(1 - s, x) +
            e^{-\pi i(1-s)/2} \zeta(1 - s, 1 - x)
        \right). \label{eq:pzhz}
    \end{equation}
    This expression requires some attention because the
    factor $\Gamma(1 - s)$ has a pole for integer $s$, canceled by a zero
    either due to the exponentials or to the symmetry property
    \begin{equation}
        \zeta(1 - s, x) = (-1)^s \zeta(1 - s, 1 - x),
        \qquad x \in [0, 1],
        \quad s \in \mathbb N_0,
        \label{eq:hzsym}
    \end{equation}
    derivable from \cite[eq.~25.11.14]{dlmf} and the analogous symmetry of the
    Bernoulli polynomials. Thus, for $s$ close to an integer, these zeros must
    be computed accurately to all significant digits.

    In the following, I discuss the calculation of the real part of $F(x, s)$.
    The procedure for the imaginary part is similar. I implemented this
    algorithm in the open-source software
    \texttt{lsqfitgp}\footnote{\url{https://github.com/Gattocrucco/lsqfitgp}}
    and thoroughly measured its accuracy, which is 110~ULP at worst (relative
    to the maximum). See \cite{crandall2012} for possible alternatives.

    First, consider the real part of \autoref{eq:pzhz}
    \begin{equation}
        \Re F(x, s) = \frac{\Gamma(1 - s)}{(2\pi)^{1-s}}
        \sin\left( \frac\pi2 s \right)
        \big( \zeta(1 - s, x) + \zeta(1 - s, 1 - x) \big).
    \end{equation}
    For even $s$, the zero is in the sine term. The standard libraries can
    handle accurately the computation if $s$ is near zero. For other values, it
    is necessary to take the difference between $s$ and its nearest even
    integer, and change the sign of the sine appropriately.
    
    For odd $s$, the zero is in the sum of Hurwitz zetas, thus it must be
    computed with full accuracy for almost odd $s$. I use the equations
    (\cite[eq.~25.11.3 and 25.11.10]{dlmf})
    \begin{align}
        \zeta(s, a) &= \zeta(s, a + 1) + a^{-s}, \label{eq:hztr} \\
        \zeta(s, a) &= \sum_{n=0}^\infty
        \frac {(s)_n} {n!} \zeta(n + s) (1 - a)^n, \\
        &\qquad |1 - a| < 1,
        \quad (s)_n = s(s+1)\cdots(s+n-1), \notag
    \end{align}
    to write the sum as
    \begin{equation}
        \zeta(1 - s, x) + \zeta(1 - s, 1 - x) =
        x^{s-1} +
        2 \sum_{\text{even $n=0$}}^\infty
        \frac {(1-s)_n} {n!} \zeta(n + 1 - s) x^n. \label{eq:sumhz}
    \end{equation}
    Due to the symmetries of $F(x, s)$, I can take $x \in [0, 1/2]$, and the
    term $(1-s)_n/n!$ is well bounded for $s < 10$, thus the series in
    \autoref{eq:sumhz} is geometrically convergent.
    
    Consider how the series behaves as $s$ gets close to an odd integer $m$.
    $\zeta(s) = 0$ for even negative $s$ (\cite[eq.~25.6.4]{dlmf}), thus all
    the terms tend to zero for $n < m - 1$. For $n > m - 1$, the Pochhammer
    symbol $(1-s)_n$ contains the factor $(1 - s + m - 1)$, thus all those
    terms tend to zero too. The only nonzero term is the $n = m - 1$ one, which
    for $s = m$ yields
    \begin{equation}
        \frac{n!}{n!} \zeta(0) x^{s-1} = -\frac12 x^{s-1},
    \end{equation}
    thus canceling the external power in \autoref{eq:sumhz}.
    
    The implications are that we have to compute accurately both the sum of the
    $(m-1)$-th term with $x^{s-1}$, and the zeta function near its zeroes. For
    the latter, use the reflection formula (\cite[eq.~25.4.1]{dlmf})
    \begin{equation}
        \zeta(1 - s) =
        2(2\pi)^{-s}
        \cos\left( \frac\pi2 s \right)
        \Gamma(s) \zeta(s),
    \end{equation}
    such that the zero is given by the cosine term. For the power, let $s = 1 +
    q + u$, with even integer $q$ and $|u| \le 1/2$, and write the sum as
    \begin{align}
        x^{q+u} &+ 2 \frac{(-q-u)_q}{q!} \zeta(-u) x^q = \\
        &= x^q \Bigg(
            e^{u \log x} - 1 + {} \label{eq:expm1} \\
            &\hphantom{{}=x^q \Bigg(}
            2 \frac{\Gamma(1 + q + u)}{\Gamma(1 + q)\Gamma(1 + u)}
            (\zeta(-u) - \zeta(0)) + {} \label{eq:zeta0} \\
            &\hphantom{{}=x^q \Bigg(}
            1 - \frac{\Gamma(1 + q + u)}{\Gamma(1 + q)\Gamma(1 + u)} \Bigg).
            \label{eq:gaminc}
    \end{align}
    Term \eqref{eq:expm1} can be computed with the standard function
    \texttt{expm1}. The difference in \eqref{eq:zeta0} can be computed with the
    Taylor expansion around zero of the pole-free zeta $\tilde\zeta(s) =
    \zeta(s) - 1/(s-1)$, yielding $1 + \zeta(s) = \tilde\zeta(s) + s/(s-1)$.
    Finally, I write the difference in \eqref{eq:gaminc} as
    \begin{equation}
        \frac{\Gamma(1 + q + u)}{\Gamma(1 + q)\Gamma(1 + u)} - 1 =
        e^{\log\Gamma(1 + q + u) - \log\Gamma(1 + q) - \log\Gamma(1 + u)} - 1,
    \end{equation}
    where the Taylor series of $\log\Gamma(1 + q + u) - \log\Gamma(1 + q)$ and
    $\log\Gamma(1 + u)$ can be generated with the generally available polygamma
    function $\psi_n$.
    
    For $s$ exactly an integer, since the above algorithm is accurate
    arbitrarily close to an integer, just multiply $s$ by $1 + \varepsilon$
    before the computation, $\varepsilon$ being the ULP of~1.
    
    Since the special functions $\Gamma$, $\zeta$ and $\psi_n$ need to be
    calculated on values depending only on $s$ and not $x$, this scheme scales
    well with the number of points the covariance function has to be evaluated
    at. The number of terms to be summed either in \autoref{eq:pzdef} or
    \autoref{eq:sumhz} is less than~50 for 53~bit floating point precision.
    
    \section{Conclusions}
    \label{sec:concl}
    
    I have shown how to use in practice the covariance function of a power-law
    spectrum one-dimensional periodic process. In relation to the initial
    problem addressed by Stein (see \autoref{sec:intro}), I left out two
    aspects: 1) the inverse correlation length $\alpha$, and 2) the
    multidimensional case.
    
    To change the correlation length, consider the Lerch zeta function
    (\autoref{eq:lerch}) with parameter $\alpha$ (note: not the same of Stein).
    For moderate integer $\alpha$, \cite[eq.~25.14.4]{dlmf}, allows to express
    the Lerch function in terms of the periodic zeta function described here.
    For more general algorithms, see \cite{crandall2012}; algorithm~2 in
    particular (\cite[eq.~9.555(2)]{gradshtein2014}) gives a series similar to
    the one in \autoref{eq:sumhz}, likewise requiring care with pole canceling
    in the implementation, for $\alpha \in (0, 1]$, which again can be extended
    to arbitrary but not too large $\alpha$ with repeated application of
    \cite[eq.~9.551(1-2)]{gradshtein2014}.
    
    To increase the dimensionality, I would need to efficiently evaluate the sum
    \begin{equation}
        \sum_{\substack{\mathbf n\in\mathbb Z^d \\ \mathbf n \ne \mathbf 0}}
        \frac
        {e^{2\pi i \mathbf n \cdot \mathbf x}}
        {|\mathbf n|^s},
    \end{equation}
    which I am currently not able to do. An alternative of course is the
    separable sum or product of kernels over each axis.
    
    A more practical aspect that I've neglected is the calculation of first and
    possibly second derivatives w.r.t.\ $\nu$, which would be useful for
    inference algorithms, from empirical Bayes to Markov chain Monte Carlo. I
    leave that to future work.
    
    \appendix
    
    \section{Proofs}
    \label{sec:proofs}
    
    \paragraph{Smoothness of the process}
    
    I consider continuity and differentiability in the mean-square sense
    (\cite[\S2.4]{stein1999}). A process $f(x)$ is mean-square continuous if
    \begin{equation}
        \lim_{x\to y} E\big[ (f(x) - f(y))^2 \big] = 0,
    \end{equation}
    and mean-square differentiable if there exists $f'(x)$ such that
    \begin{equation}
        \lim_{h\to 0} E\left[ \left(
            \frac {f(x + h) - f(x)} {h} - f'(x)
        \right)^2 \right] = 0.
    \end{equation}

    These expected values translate into analogous expressions for the
    covariance function $K$. In particular, a stationary process is M.S.
    continuous iff $K$ is continuous in zero, and $m$ times M.S. differentiable
    iff $K$ is $2m$ times derivable in zero.
    
    It follows that for $\nu > 0$, $Z_\nu(x)$ (\autoref{eq:znu}) induces a M.S.
    continuous process. However, visual inspection of numerical simulations of
    the process shows that for $\nu < 1/2$ it appears very discontinuous. To
    characterize this behavior, I consider Lipschitz continuity. I say that a
    process is M.S. Lipschitz-continuous if
    \begin{equation}
        \exists C \, \forall x, y :
        E\big[ (f(x) - f(y))^2 \big] \le C (x - y)^2.
    \end{equation}
    
    This property is violated if $\lim_{x\to 0^+} K_\nu'(x) = -\infty$, since
    the ratio $(K_\nu(0) - K_\nu(x-y)) / (x - y)$ becomes arbitrarily large as
    $x\to y$. To see that this is the case, consider that $\zeta(s, a)$
    diverges for $a\to 0^+$ at fixed $s > 0$ due to \autoref{eq:hztr}, and
    that, for $0 < \nu < 1/2$, \autoref{eq:pzhz} applies to $K_\nu'(x)$ with $0
    < s < 1$.\footnote{\cite[eq.~25.13.2]{dlmf} reports that the range of
    validity of \autoref{eq:pzhz} is $\Re s > 1$. However, at least for the set
    of values I am considering, that relation can be derived from eq.~25.13.3,
    which only requires $\Re s > 0$. See also eq.~25.11.17 and \cite[ex.~3
    p.~273]{apostol1976}.} This implies that $F(x, s)$ with $s \in (0, 1)$
    diverges for $x \to 0^+$, both in the real and imaginary part. But the
    imaginary part gives the derivative of $\Re F(x, s + 1) \propto K_\nu(x)$.
    
    The same continuity property can be proven about the Matérn class using
    \cite[eq.~10.29.4 and~10.27.3]{dlmf}.
    
    \paragraph{Zero $\nu$ limit}
    
    Since $F(x, 1)$ is finite for non-integer $x$ (property~\ref{it:pzconv}) and
    $\zeta(s) \sim 1/(s-1)$, $\lim_{\nu\to 0^+}Z_\nu(x) = 0$ for $x \bmod 1 \ne
    0$.
    
    A similar limit can be proven for the Matérn class $M_\nu(x)$
    (\autoref{eq:matern}). The normalization is chosen to have $M_\nu(0) = 1$,
    so it remains to show that $\lim_{\nu\to 0^+} M_\nu(x) = 0$ for $x > 0$. If
    it wasn't for the $\sqrt{2\nu}$ rescaling, $x^\nu K_\nu(x)/\Gamma(\nu)$
    would trivially converge to zero due to the finiteness of $K_0(x)$ and the
    pole of $\Gamma$. However, the $\sqrt\nu$ factor brings the evaluation
    point closer to zero, where $K_\nu$ diverges. Consider the series expansion
    (\cite[eq.~10.27.4, 10.25.2]{dlmf})
    \begin{equation}
        K_\nu(2z) = \frac\pi2 \sum_{k=0}^\infty
        \frac 1 {k! \sin(\pi\nu)} \left(
            \frac {z^{-\nu}} {\Gamma(1 + k - \nu)} -
            \frac {z^\nu} {\Gamma(1 + k + \nu)}
        \right) z^{2k}.
    \end{equation}
    Replacing $z \mapsto \sqrt\nu x$, the leading term in $\nu$ is the first
    one, which goes like
    \begin{equation}
        (\sqrt\nu x)^{-\nu}-(\sqrt\nu x)^\nu =
        -2\sinh\left(\frac12 \nu\log\nu + \nu\log x\right) \sim
        \nu\log\nu \to 0.
    \end{equation}
    The $\sin(\pi\nu) \sim \nu$ denominator is cancelled by the $\Gamma(\nu)
    \sim 1/\nu$ normalization.
    
    \paragraph{Positive definiteness on the 3-sphere}
    
    \cite[corollary~4a]{gneiting2013}, gives the following necessary and
    sufficient condition for positive definiteness on $\mathbb S^3$ of a
    function $\psi$ defined in terms of the geodesic distance. Let
    \begin{equation}
        \psi(\theta) = \sum_{k=0}^\infty b_n \cos(n\theta),
        \qquad \theta \in [0, \pi],
    \end{equation}
    be the Fourier series expansion of $\psi$. Then $\psi$ is a valid
    covariance function on $\mathbb S^3$, with $\theta$ the great circle
    distance, i.e., the angular length of the shorter arc connecting two
    points, or equivalently the arc along the intersection of a plane passing
    by the center and the two points with the sphere, if and only if
    \begin{enumerate}
        \item $b_2 \le 2 b_0$, and \label{it:initial}
        \item $b_{n+2} \le b_n$ for $n \ge 1$. \label{it:lag2}
    \end{enumerate}
    
    The series of $F(x, s)$ (\autoref{eq:pzdef}) of course satisfies
    condition~\ref{it:lag2}. To also have~\ref{it:initial}, it is necessary
    to add a constant term $b_0$ such that $2b_0 \ge 2^{-s}$. This leads to
    the expression given at the end of \autoref{sec:pz}.
    
    \printbibliography
    
\end{document}